\documentclass[
twocolumn,%
nofootinbib,%
superscriptaddress,%
amsfonts,amsmath]{revtex4}

\usepackage{amsmath}

\begin{document}

\title{Stability analysis of black holes in massive gravity: a unified treatment}

\author{Eugeny~Babichev} 
\affiliation{Laboratoire de Physique Th\'eorique d'Orsay,
B\^atiment 210, Universit\'e Paris-Sud 11,
F-91405 Orsay Cedex, France}

\author{Alessandro Fabbri}
\affiliation{Museo Storico della Fisica e Centro Studi e Ricerche ’Enrico Fermi’, Piazza del Viminale 1, 00184 Roma, Italy; Dipartimento di Fisica dell’Universita` di Bologna,
Via Irnerio 46, 40126 Bologna, Italy; Departamento de F\'isica Te\'orica and IFIC, Universidad de Valencia-CSIC, C. Dr. Moliner 50, 46100 Burjassot, Spain}
\affiliation{Laboratoire de Physique Th\'eorique d'Orsay,
B\^atiment 210, Universit\'e Paris-Sud 11,
F-91405 Orsay Cedex, France}

\begin{abstract}
We consider the analytic solutions of massive (bi)gravity which can be written in a simple form using advanced Eddington-Finkelstein coordinates.
We analyse the stability of these solutions against radial perturbations.
First we recover the previously obtained result on the instability of the bidiagonal bi-Schwarzschild solutions. 
In the non-bidiagonal case (which contains, in particular, the Schwarzschild solution with Minkowski fiducial metric) 
we show that generically there are physical spherically symmetric perturbations, but no unstable modes.

\end{abstract}

\date{\today}


\maketitle

General Relativity (GR) theory is very successful in the explanation of various gravity phenomena.
In particular, GR passes the local gravity tests, which give the most impressive constraints. 
In spite of this obvious success, there are still unresolved puzzles, e.g. Dark Energy and Dark Matter, which 
motivate to search ways to modify GR. 
One way to modify gravity follows an idea of Fierz and Pauli to equip the graviton with a non-zero mass~\cite{Fierz:1939ix}.
However, a close examination of the massive gravity models revealed pathologies, in particular,
the van~Dam-Veltman-Zakharov discontinuity~\cite{vanDam:1970vg} as well as the presence of a ghost in the spectrum~\cite{Boulware:1973my}.
The solution to the first problem has been conjectured by Vainshtein~\cite{Vainshtein:1972sx} and confirmed much later in~\cite{Babichev:2009us} (for a recent review see, e.g.~\cite{Babichev:2013usa}).
On the other hand, the latter problem (related to the presence of the Ostrogradski ghost~\cite{Ostrogradski}) 
has been recently addressed in a series of works~\cite{deRham:2010ik}, where it was shown that for a subclass of  massive potentials
the Ostrogradski ghost does not appear, both in the model with one and two dynamical 
metrics~\footnote{The absence of this ghost degree of freedom does not necessarily mean though that all other degrees of freedom are healthy.}.
Other possible problems have been also discussed in the literature~\cite{ArkaniHamed:2002sp,Deser:2012qx}. 
For a review on massive gravity see~\cite{deRham:2014zqa}.

Black hole solutions in massive gravity are of two types~\cite{VolkovTasinato}. 
In the first, both the physical ($g$) and fiducial ($f$) metrics are diagonal (bi-diagonal case). 
In the second $g$ and $f$ are not simultaneously diagonal (non bi-diagonal case).

For the bi-diagonal case,
it was shown in   \cite{Deffayet:2011rh} that regularity at the horizon requires that, for spherically symmetric solutions, the two metrics must share the same Killing horizon. 
This condition is straightforwardly satisfied when we take $f=g$, in which case the massive interaction term between the two metrics vanishes and the equations of motion reduce to those of GR. In \cite{Babichev:2013una} we considered the two metrics to be equal to the same Schwarzschild solution (bi-Schwarzschild solution) and showed that, despite the `triviality' of the background, the massive interaction term shows up nontrivially in the linear perturbations featuring the Gregory-Laflamme (GL) instability \cite{Gregory:1993vy} of higher-dimensional black strings. This is however a mild instability, since for the physically meaningful case where the
graviton mass is $m\sim \frac{1}{H}$, where $H$ is the Hubble scale, 
the timescale of the instability $\tau^{GL} \sim \frac{1}{m}$ is of the order of the Hubble time.
This result is valid for both one ($g$) or two ($g$ and $f$) dynamical metrics and, generically, when the two Schwarzschild metrics are proportional with a constant conformal factor. 
The existence of such instability was confirmed numerically in \cite{Brito:2013wya}. Subsequent investigations by the same authors found possible candidates (typically, in the $m\sim \frac{1}{r_S}$ regime, where $r_S$ is the Schwarzschild radius) for the end-point evolution \cite{Brito:2013xaa}. 

It is important to understand whether or  not the presence of this instability is a generic feature of (physically relevant) static massive gravity black holes. For this reason, in this Letter we shall extend our perturbation analysis to the technically more involved case of non bi-diagonal black hole solutions. 
Another motivation of this work is that, as already mentioned, in the bi-diagonal case regularity conditions \cite{Deffayet:2011rh} do not allow the fiducial metric $f$ (fixed or dynamical) to be flat (Minkowski) 
and this might seem unnatural.  In the non bi-diagonal case both metrics can be nicely expressed in ingoing Eddington-Finkelstein (EF) coordinates, a choice that 
satisfies automatically the regularity condition on the future horizon~\footnote{As it will become clear below this coordinate system greatly simplifies the perturbation analysis. 
For the solutions that cannot be written in this form the analysis is much more involved.}. 
We shall write down solutions that contain the most interesting Schwarzschild (g) -- Minkowski (f) case. Remarkably, the s-wave perturbation equations can be solved analytically.
We find, generically, a nontrivial massive perturbation term but, unlike the bi-diagonal bi-Schwarzschild case, no unstable modes.

The action for the dRGT (bi-gravity) model can be written as follows~\cite{deRham:2010ik}
\begin{eqnarray}\label{action1}
S &=&  M^2_P\int d^4 x \sqrt{-g} \left(\frac{R[g]}{2} + m^2 {\cal U} [g,f] - m^2 \Lambda_g  \right) 	+
 S_m [g] \nonumber \\ &+& \frac{\kappa M^2_P}{2}\int d^4 x \sqrt{-f} \left(\mathcal{R}[f] - m^2 \Lambda_f \right).\end{eqnarray}
The interaction potential ${\cal U}[g,f]$ is expressed in terms of the matrix $\mathcal{K}^\mu_\nu = \delta^\mu_\nu - \gamma^\mu_\nu$, 
where $\gamma^\mu_\nu = \sqrt{g^{\mu\alpha}f_{\alpha\nu}} $. 
The potential $\cal{U}$ consists of three pieces, 
$
	\mathcal{U} \equiv  \mathcal{U}_2 + \alpha_3 \mathcal{U}_3 + \alpha_4 \mathcal{U}_4,
$
with $\alpha_3$ and $\alpha_4$ being parameters of the theory, and each of them reads,
\begin{equation}
\label{UdRGT}
\begin{aligned}
	\mathcal{U}_2 &= \frac{1}{2!}\left( [\mathcal{K}]^2 - [\mathcal{K}^2] \right), \\
	\mathcal{U}_3 &= \frac{1}{3!}\left( [\mathcal{K}]^3 -3 [\mathcal{K}] [\mathcal{K}^2] +2[\mathcal{K}^3]\right), \,	
	\mathcal{U}_4 = \det (\mathcal{K}),
\end{aligned}
\end{equation}
where  $[\mathcal{K}]\equiv \mathcal{K}^\rho_\rho$ and 
	$[\mathcal{K}^n]\equiv  (\mathcal{K}^n)^\rho_\rho $.

The variation of the action with respect to $g$ and $f$ in the vacuum gives
\begin{eqnarray}
	G^{\mu}_{\phantom{\mu}\nu}  &=&  m^2\left( T^{\mu}_{\phantom{\mu}\nu} -  \Lambda_g \delta^\mu_\nu \right),\label{Eg}\\ 	
	\mathcal{G}^{\mu}_{\phantom{\mu}\nu} & = & m^2 \left(
		 \frac{\sqrt{-g}}{\sqrt{-f}}\frac{ \mathcal{T}^{\mu}_{\phantom{\mu}\nu}}{\kappa} - \Lambda_f \delta^\mu_\nu \right) ,\label{Ef}
\end{eqnarray}
where $G^{\mu}_{\phantom{\mu}\nu}$ and $\mathcal{G}^{\mu}_{\phantom{\mu}\nu}$ are the corresponding Einstein tensors for the two  metrics $g$ and $f$, 
$T^{\mu}_{\phantom{\mu}\nu} \equiv \mathcal{U} \delta^{\mu}_{\nu} - 2 g^{\mu\alpha}\frac{\delta \mathcal{U}}{\delta g^{\nu\alpha}}$
and $\mathcal{T}^\mu_{\phantom{\mu}\nu} = -T^\mu_{\phantom{\mu}\nu} + \mathcal{U}\delta^\mu_\nu$.
When $f$ is not dynamical then Eq.~(\ref{Ef}) is absent.

We shall write down the metric solutions in the bi-advanced Eddington-Finkelstein (biEF) form
\begin{eqnarray}\label{sol}
ds_g^2 & = & -\left(1-\frac{r_g}{r}\right)dv^2 +2dvdr+r^2 d\Omega^2,\label{metricg}\\
ds_f^2 & = & C^2\left[-\left(1-\frac{r_f}r \right)dv^2 +2dvdr+r^2 d\Omega^2\right], \label{metricf}
\end{eqnarray}
where $C$ is a constant (conformal factor) and $r_g$ and $r_f$ are the two (in general different) Schwarzschild radii for the two metrics.
The only non-diagonal terms of $T^{\mu}_{\phantom{\mu}\nu}$ and $\mathcal{T}^\mu_{\phantom{\mu}\nu}$ read,
\begin{equation}\label{Toff}
	T^r_{\phantom{r}v} =-\mathcal{T}^r_{\phantom{r}v}= \frac{C \left(\beta  (C-1)^2-2 \alpha  (C-1)+1\right) \left(r_f-r_g\right)}{2 r},
\end{equation}
where we defined $\alpha\equiv 1+\alpha_3,\ \beta \equiv \alpha_3+\alpha_4$. 
These off-diagonal terms must vanish due to the choice of the metrics 
(\ref{metricg}) and (\ref{metricf}).
This implies either $r_S=r_f$, which is equivalent to the (bi-diagonal) bi-Schwarzschild solution analysed in \cite{Babichev:2013una}, or 
\begin{equation}
\label{relation} 
	\beta  (C-1)^2-2 \alpha (C-1)+1 = 0,
\end{equation} that we will consider in detail below. The case with a flat (Minkowski) fiducial metric ($f$) falls within this class of solutions.  
For all the solutions we need to fine tune the two cosmological
constants to 
$\Lambda_g = -(C-1)(\beta(C-1)^2-3\alpha(C-1)+3),
	\Lambda_f = \frac{1}{\kappa C^3}\left( C^3 (1-\alpha+\beta) -3C^2 \beta+ 3C (\alpha+\beta)-2\alpha-\beta -1\right) $
to balance the corresponding ($\sim\delta^{\mu}_{\nu}$) contributions coming from 
$T^\mu_{\phantom{\mu}\nu}$ and $\mathcal{T}^\mu_{\phantom{\mu}\nu} $. 

Let us now consider the linear perturbations around the solutions (\ref{metricg}), (\ref{metricf}). The metric perturbations $h_{\mu\nu}^{(g)}$ and
$h_{\mu\nu}^{(f)}$ satisfy the linearised equations 
\begin{equation}\label{perteqs}
	\delta G^{\mu}_{\phantom{\mu}\nu}  =  
	m^2 \delta T^{\mu}_{\phantom{\mu}\nu}, \ \ \ 
	\delta \mathcal{G}^{\mu}_{\phantom{\mu}\nu}  =  \frac{m^2}\kappa \delta\left(
		 \frac{\sqrt{-g}}{\sqrt{-f}} \mathcal{T}^{\mu}_{\phantom{\mu}\nu}  \right)\ .
\end{equation} 
We shall look for unstable ($\Omega>0$) spherically-symmetric modes of the form 
\begin{equation}\label{swave}
h^{\mu\nu}_{(g)}=e^{\Omega v}\left(
  \begin{array}{cccc}
   h^{vv}_{(g)}(r) & h^{vr}_{(g)}(r)  & 0 & 0 \\
   h^{vr}_{(g)}(r) & h^{rr}_{(g)}(r)  & 0 & 0 \\
    0 & 0 &  \frac{h^{\theta\theta}_{(g)}(r)}{r^2} & 0 \\
   0 & 0 & 0 & \frac{h^{\theta\theta}_{(g)}(r)}{r^2 \sin^2\theta} \\
  \end{array}\right)\ ,
  \end{equation}
  along with a similar expression for $h^{\mu\nu}_{(f)}$  but with an overall $1/C^2$ term, which are regular at the future horizon and vanish asymptotically. 
  As the advanced time $v$ is regular at the future horizon, we require the metric perturbations $h^{\mu\nu}_{(g,f)}(r)$  to be regular at $r=r_g$ (and $r=r_f$). At infinity, instead, it is more suitable to use the Schwarzschild time $t \sim v-r$ to separate between temporal and spatial components. 
  Therefore, in the asymptotic region $h^{\mu\nu}_{(g,f)}(r)$  must behave as $o\left( e^{-\Omega r}\right)$ to be physically acceptable.

In the non-bidiagonal case (\ref{relation}), it turns out that the massive terms in the perturbations equations take the remarkably simple form
\begin{equation}\label{nbd}
\delta T^{\mu}_{\phantom{\mu}\nu} =
\frac{\mathcal{A} \left(r_S-r_f\right) }{4 r}\, e^{\Omega v} 
\left(
\begin{array}{cccc}
 0 & 0 & 0 & 0 \\
 h_{(-)}^{\theta \theta } & 0 & 0 & 0 \\
 0 & 0 & \frac{h_{(-)}^{vv}}{2} & 0 \\
 0 & 0 & 0 & \frac{h_{(-)}^{vv}}{2}
\end{array}
\right),
\end{equation}
where 
\begin{equation}\label{AA}
\mathcal{A} = \frac{C^2\left(\beta(C-1)^2-1\right)}{C-1},
\end{equation}
and $\delta\left( \frac{\sqrt{-g}}{\sqrt{-f}} \mathcal{T}^{\mu}_{\phantom{\mu}\nu}\right)=-\delta T^\mu_{\phantom{\mu}\nu} $, 
$h^{\mu\nu}_{(-)} \equiv h^{\mu\nu} _{(g)}- C^2h^{\mu\nu}_{(f)}$ 
[So that, e.g. $h^{vv}_{(-)}(r)=h^{vv}_{(g)}(r)-h^{vv}_{(f)}(r)$, taking into account the factor $1/C^2$ in the definition of $h^{\mu\nu}_{(f)}$].
This is to be compared with the Pauli-Fierz-like form in the  bi-diagonal (bi-Schwarzschild) case 
\begin{equation}\label{bd}
\begin{aligned}
\delta T^\mu_{\phantom{\mu}\nu}=\frac{C}{2}\left(\beta(C-1)^2-2\alpha(C-1)+1\right) \\
\times e^{\Omega v} \left(\delta^{\mu}_\nu h^{(-)}-h^{\mu (-)}_{\phantom{\mu}\nu}\right),
\end{aligned}
\end{equation}
together with, again, $\delta\left( \frac{\sqrt{-g}}{\sqrt{-f}} \mathcal{T}^{\mu}_{\phantom{\mu}\nu}\right)=-\delta T^\mu_{\phantom{\mu}\nu} $.   
It is interesting to note that at the intersection of the two branches of solutions, $r_S=r_f$ in (\ref{nbd}) and $C$ fixed by  (\ref{relation}) in (\ref{bd}), we get  $\delta T^{\mu}_{\phantom{\mu}\nu}=0$, i.e. the perturbation equations are those of GR. This is not true for  $r_S\neq r_f$, and in particular when the fiducial metric $f$ is flat ($r_f=0$), unless $\mathcal{A}=0$ 
in (\ref{nbd}), i.e. 
$\beta=(C-1)^{-2}$ implying, from (\ref{relation}), $\beta=\alpha^2$. This choice, that was also made in \cite{Kodama:2013rea}, is not generic~\footnote{The ``degenerate'' case $\beta=\alpha^2$ seems to be special since it allows for more black hole solutions~\cite{Volkov:2012wp}.}.

Taking the divergence of Eq.~(\ref{perteqs}) and using the Bianchi identities for the Einstein tensors 
we obtain the constraint 
$ \nabla^\nu_{(f)}\delta \left(\frac{\sqrt{-g}}{\sqrt{-f}}\mathcal{T}^{\mu}_{\phantom{\mu}\nu}\right)\propto \nabla^\nu_{(g)}\delta T^{\mu}_{\phantom{\mu}\nu} =0$ which 
in non-bidiagonal case (\ref{nbd}) reads
\begin{equation}\label{divg}
\frac{\mathcal{A} \left(r_g-r_f\right) }{4 r^2}\, e^{\Omega v}
\left\{-\left( r h_{(-)}^{\theta \theta}\right)', h_{(-)}^{vv},0,0\right\}  =0.
\end{equation}
From the above equation one  immediately finds the conditions
\begin{equation}\label{constnbd} 
h^{vv}_{(-)} =0 ,\; h^{\theta\theta}_{(-)} = \frac{c_0}r,
\end{equation} 
where $c_0$ is an integration constant. 
Plugging this back to (\ref{nbd}) one can see that, generically, there is only one nontrivial (off-diagonal) component
of the matrix $\delta T^{\mu}_{\phantom{\mu}\nu}$ proportional to $h_{(-)}^{\theta \theta }$. 
This is very different from the bi-diagonal case (\ref{bd}), in which the constraints require $h^{\mu\nu}_{(-)}$ to be divergence free and traceless,
   \begin{equation}\label{constbd}
   \nabla_{\mu}h^{\mu\nu}_{(-)}=h_{(-)}=0\ .
   \end{equation}   
 
 Let us now discuss the solution to the perturbation equations. 
 In the bi-diagonal (bi-Schwarzschild) case subtraction of the two Eqs. (\ref{perteqs})  subject to the constraints (\ref{constbd}) leads to the massive Lichnerowicz equation for the metric component $h^{\mu\nu}_{(-)}$,
 \begin{equation}\label{GL}
	 \Box h_{(-)}^{\mu\nu} + 2 R_{\sigma\mu\lambda\nu} h_{(-)}^{\lambda\sigma} = m_{eff}^2 h_{(-)}^{\mu\nu}\ ,
 \end{equation}
where 
\begin{equation*}
	m_{eff}^2=\frac{m^2}{2}\left(1+\frac{1}{\kappa}\right)C\left(\beta(C-1)^2-2\alpha(C-1)+1\right),
\end{equation*}
which is known to exhibit  GL  unstable modes for $0<m'<O(1/r_S)$~\cite{Gregory:1993vy}. 
Note that there is a  region in parameter space 
for which $m_{eff}^2$ becomes negative, and hence 
Eq.~(\ref{GL}) does not correspond to the case studied in~\cite{Gregory:1993vy}. 
The flip of sign of $m_{eff}^2$, however, signals that the spin-0 part of the graviton becomes a ghost, therefore the vacuum is unstable at the quantum level.
This can also be seen from the Higuchi bound~\cite{Higuchi:1986py}, when the de-Sitter curvature goes to zero.

In non bi-diagonal case we cannot  single out the gauge-invariant (massive) metric component in the same way by considering linear combinations of Eqs.~(\ref{perteqs}). 
Nevertheless, the simple form of the massive matrix (\ref{nbd}) together with the constraints (\ref{constnbd}) lead  to an analytical resolution of the perturbation equations.
General solutions for $h^{\mu\nu}_{(g)}$ and $h^{\mu\nu}_{(f)}$ contain a part that is gauge-dependent (same as in GR) plus a particular solution to the full equations, i.e.  
\begin{equation}
	h^{\mu\nu}_{(g,f)} =  h^{\mu\nu (g,f)}_{GR} + h^{\mu\nu (g,f)}_{(m)} . 
\end{equation}
The particular (gauge-invariant) solution  is  given by a single nonzero component for each metric perturbation
\begin{eqnarray}
 	h^{rr(g)}_{(m)} &=&  \frac{\mathcal{A}(r_g-r_f) e^{ \Omega v}}{4\Omega } m^2  h_{(-)}^{\theta \theta }, \label{partsolg}\\
 	h^{rr(f)}_{(m)} &=& -\frac{h^{rr(g)}_{(m)}}{\kappa} . \label{partsolf}
 \end{eqnarray}
Both  $h^{\mu\nu (g)}_{GR}$ and $h^{\mu\nu (f)}_{GR}$ are, individually, pure gauge. 
This means that one can write them formally as  
$ h^{\mu\nu}_{GR} =  -\nabla^\mu \xi^\nu -\nabla^\nu\xi^\mu$, 
where, to recover the form~(\ref{swave}), one requires $\xi^\mu  = e^{\Omega v}\left\{ \xi^0(r),\xi^1(r),0,0 \right\}$. 
In general there are two different $\xi^\mu_{(g)}$ and $\xi^\mu_{(f)}$, and their relation is fixed by the constraint (\ref{constnbd}), namely, 
$\xi^0_{(f)} = \xi^0_{(g)} + c_1$, $\xi^1_{(f)} = \xi^1_{(g)} + \frac{c_0}2$ with $c_1$ being  the other integration constant. 
We can now use the coordinate transformation, e.g., to  completely  eliminate the perturbation of the metic $f$, i.e. 
\begin{equation}\label{hf0}
h^{\mu\nu (f)}_{GR} =0, 
\end{equation}
leaving though the non-zero $h^{\mu\nu (g)}_{GR}$
\begin{equation}\label{resexp}
h^{\mu\nu (g)}_{GR}=e^{\Omega v}
\left(
\begin{array}{cccc}
0 & \Omega  c_1 & 0 & 0 \\
 \Omega  c_1 & c_0 \left(\Omega - \frac{r_g}{2 r^2} \right) & 0 & 0 \\
 0 & 0 & c_0 r^{-3} & 0 \\
 0 & 0 & 0 & c_0 \csc ^2(\theta ) r^{-3}
\end{array}
\right).
\end{equation} 
This completes the derivation of the full solution to Eqs.~(\ref{perteqs}). 
We immediately see that (\ref{partsolg}), (\ref{partsolf}) and (\ref{resexp}) are regular at the horizon, but not at infinity~\footnote{In the bi-diagonal case the GL unstable mode is 
regular at the horizon and decays asymptotically as 
$h^{\mu\nu}_{(-)}\sim e^{ -\left(\Omega+\sqrt{\Omega^2+m_{eff}^2}\right)r } $ and thus is physically acceptable.}.
Therefore the non-bidiagonal black holes (\ref{sol}), (\ref{metricf}), (\ref{relation}) do not have unstable modes. 
The physical perturbations (\ref{partsolg}), (\ref{partsolf}) are regular when $\Omega=iw$, in which case they describe ``ingoing'' waves. 
The existence of physical static perturbations ($\Omega=0$) seems to be excluded due to the term $\Omega^{-1}$ in (\ref{partsolg}), (\ref{partsolf})
unless we take $c_0\sim \Omega$, in which case $h^{\mu\nu (f)}_{GR}$ vanishes and 
the only nonvanishing contribution left in $h^{rr}$ is~$\sim 1/r$, which describes the same solutions (\ref{metricg}), (\ref{metricf}) with, however, different $r_g$ and $r_f$.
Therefore we exclude the existence of other branch of solutions close to this family. 

Having the result~(\ref{resexp}), which is valid for the case of two dynamical metrics, 
it is easy to get to the case of one dynamical metric, in particular when the fiducial metric is Minkowski (as in the original dRGT model).
In fact, Eqs.~(\ref{partsolg}) and (\ref{resexp}) is the solution in this case as well, with the metric $f$ unperturbed, $h^{\mu\nu(f)} = 0$. 

It is interesting to point out that the non-bidiagonal black hole solutions require the specific choice of the conformal factor $C$, given by~(\ref{relation}).
At the same time such a choice of $C$ implies no scalar mode (i.e. no s-wave) in the case of bi-flat spacetime, 
i.e. the helicity-0 mode is (infinitely) strongly coupled, see e.g.~\cite{Comelli:2011wq}.
This can be seen directly from our analysis: going to spatial infinity, $r\to\infty$, the ``massive'' part of the  ($\Omega=iw$) perturbations $h^{\mu\nu(g,f)}_{(m)}$ disappears as it can be seen 
from~(\ref{partsolg}), (\ref{partsolf}) and  (\ref{constnbd}).
The presence of curvature restores the scalar degree of freedom, giving a non-trivial
solution, which cannot be gauged away.

To summarize, we presented a stability analysis of both bidiagonal and non-bidiagonal black hole solutions against spherically symmetric perturbations. 
We considered the solutions of massive gravity which can be written in the simple form~(\ref{sol}) and (\ref{metricf}) using advanced Eddington-Finkelstein coordinates.
We then studied the perturbations of the metric(s) around these solutions.
We confirmed our previous result on the instability of bi-diagonal solutions~\cite{Babichev:2013una}.
On the other hand, we found that non-bidiagonal solutions generically possess physical s-wave perturbations, Eqs.~(\ref{partsolg}), (\ref{partsolf}) and (\ref{resexp}),
however, there are no unstable spherically symmetric modes. 
For a particular choice of the parameters of the model (namely, $\beta=\alpha^2$) 
the mass term for in the perturbation equations  of the non-bidiagonal solutions is identically zero,
and one obtains the same perturbation equation(s) as in GR. 
There are several open questions, which go beyond the scope of this Letter.
In particular, we  focussed on spherically symmetric perturbations, while it is important to extend our analysis to non-spherical perturbations as well.
Also we have not investigated in detail the ghost issue (e.g. for the non-bidiagonal case).
These questions are left for future work.

{\em Acknowledgments}. We thank Marco Crisostomi and Mikhail Volkov for useful discussions.
 The work of E.B. was supported in part by the Grant No. RFBR 13-02-00257-a.

\end{document}